\documentclass[
 reprint,
 superscriptaddress,
 amsmath,
 amssymb,
 aps,
 longbibliography
]{revtex4-1}

\usepackage{graphicx}  % Include figure files
\usepackage{dcolumn}   % Align table columns on decimal point
\usepackage{bm}        % Bold math
\usepackage{amstext,amsthm}
\usepackage{braket}
\usepackage{bbold}
\usepackage{bbm}
\usepackage{xcolor}
\usepackage{gensymb}
\usepackage{soul}

\usepackage{hyperref}% standard LaTeX stuff

\hypersetup{colorlinks,linkcolor=blue,citecolor=blue,urlcolor=blue}

\usepackage{etoolbox}

\AtBeginEnvironment{equation}{\vspace{6pt}}

\begin{document}

\setlength{\abovedisplayskip}{1pt}

\title{Probing Single-Particle Spatial Extent With Helical Neutron Wavefronts}
\author{D. Sarenac}
\email{dusansar@buffalo.edu}

\affiliation{Department of Physics, University at Buffalo, State University of New York, Buffalo, New York 14260, USA}

\author{O. Lailey} 
\affiliation{Institute for Quantum Computing, University of Waterloo,  Waterloo, ON, Canada, N2L3G1}
\affiliation{Department of Physics and Astronomy, University of Waterloo, Waterloo, ON, Canada, N2L3G1}

\author{C. W. Clark}
\affiliation{Joint Quantum Institute, National Institute of Standards and Technology and University of Maryland, College Park, Maryland 20742, USA}
\author{D. G. Cory}
\affiliation{Institute for Quantum Computing, University of Waterloo,  Waterloo, ON, Canada, N2L3G1}
\affiliation{Department of Chemistry, University of Waterloo, Waterloo, ON, Canada, N2L3G1}

\author{H. Ekinci} 
\affiliation{Institute for Quantum Computing, University of Waterloo,  Waterloo, ON, Canada, N2L3G1}
\affiliation{Department of Physics and Astronomy, University of Waterloo, Waterloo, ON, Canada, N2L3G1}

\author{D. V. Garrad}
\affiliation{Department of Physics, University of Waterloo, Waterloo, ON, Canada, N2L3G1}

\author{M. G. Huber}
\affiliation{National Institute of Standards and Technology, Gaithersburg, Maryland 20899, USA}

\author{L. Matthews}
\affiliation{ISIS Neutron and Muon Source, Rutherford Appleton Laboratory, Harwell Oxford, Didcot, UK}

\author{N. Shentevski}

\affiliation{Department of Physics, University at Buffalo, State University of New York, Buffalo, New York 14260, USA}

\author{P. R. Vadnere}

\affiliation{Department of Physics, University at Buffalo, State University of New York, Buffalo, New York 14260, USA}

\author{D. A. Pushin}
\email{dmitry.pushin@uwaterloo.ca}
\affiliation{Institute for Quantum Computing, University of Waterloo,  Waterloo, ON, Canada, N2L3G1}
\affiliation{Department of Physics and Astronomy, University of Waterloo, Waterloo, ON, Canada, N2L3G1}

\begin{abstract}

Distinguishing transverse coherence length from single-particle wavepacket extent is fundamentally challenging, as both manifest through spatial broadening of observed intensity profiles in conventional experiments. Here we introduce a method based on helical neutron wavefronts that enables this separation. Helical neutron states produce annular intensity profiles whose peak radius depends on the transverse wavepacket extent, while coherence length only contributes to profile broadening. In our experimental geometry we measure a beam divergence of $\approx1.1~\mathrm{mrad}$, corresponding to a transverse coherence length of $\approx 180~\mathrm{nm}$. In contrast, the same measurement places a lower bound of $\geq 2~\mu\mathrm{m}$ on the spatial extent of the individual neutron wavepackets, more than an order of magnitude larger than the coherence length. These results provide direct experimental evidence that transverse coherence length and single-particle wavepacket extent are distinct physical quantities, resolving a longstanding source of confusion in the neutron literature.
\end{abstract}
\maketitle

%\section{Introduction}
\section{Introduction}

Neutron beams are foundational tools in probing the structure and dynamics of matter across physics, materials science, and biology~\cite{rauch2015neutron, kardjilov2011neutron, lindner2024neutrons}. In neutron experiments, the coherence of the beam sets a fundamental limit on the spatial correlations and interaction scales that can be meaningfully resolved~\cite{utsuro2010handbook,fischer2006neutron}. The transverse coherence length, which is inversely proportional to the angular spread of the beam, is typically measured to be on the nanometer to micron scale~\cite{kaiser1983, klein1983longitudinal, pushin2008measurements, Rauch1996}. The distinction between transverse coherence length and the spatial extent of a single neutron’s wavefunction has not always been made explicit in the neutron literature, with related discussions appearing in earliest works~\cite{shull1969single}. However, the explicit identification of coherence length with the spatial extent of a single neutron’s wavefunction in several recent studies~\cite{cappelletti2018intrinsic,treimer2024computerized,treimer2025neutron} has led to conceptual misinterpretations and, for example, predictions of orbital angular momentum (OAM) dependent cross sections~\cite{jach2022method} that are inconsistent with experimental observations~\cite{sarenac2026experimental}. A more accurate description instead follows from applying Glauber’s framework for partially coherent fields~\cite{glauber1963quantum,glauber1963coherent}. In this picture, neutron beams are not composed of small localized packets, but are better understood as ensembles of extended wavepackets whose mean propagation directions are distributed over a range of angles.

The spatial extent of an individual neutron wavepacket determines the region over which a single neutron can interact coherently with its environment, yet it has proven difficult to determine experimentally because conventional measurements are primarily sensitive to ensemble coherence. Since both transverse coherence length and wavepacket extent manifest as spatial broadening of measured intensity distributions, most neutron experiments access only the coherence properties of the ensemble, with only limited experimental attempts to distinguish the two~\cite{majkrzak2014determination,majkrzak2022effect,Hamilton1983}.

Here we address this problem using helical neutron wavefronts carrying well defined OAM~\cite{sarenac2022experimental,sarenac2024small,lailey2025multimode}. We show that these states provide an observable that depends directly on the transverse wavepacket extent while remaining largely insensitive to coherence-induced broadening. In the present work, we find a transverse coherence length of approximately $180~\mathrm{nm}$, while simultaneously establishing a lower bound of $\geq 2~\mu\mathrm{m}$ on the extent of the individual neutron wavepackets. These results establish the experimental distinction between beam coherence and single-particle wavepacket extent. Within this framework, post-selection through apertures and collimation optics can modify or reveal the coherence properties of the beam by selecting subsets of the ensemble with narrower angular distributions. The resulting increase in measured coherence length should not be interpreted as a corresponding increase in the spatial extent of the individual neutron wavepackets, since post-selection acts on ensemble correlations rather than on the intrinsic extent of the single-particle state. This asymmetry highlights the fundamentally different physical origins of beam coherence length and single particle wavepacket extent.  

\subsection{Beam Divergence and Transverse Coherence Length}

Neutron beams emerging from reactor moderators are well modeled as thermal sources with random phase relationships across the emitting surface. As a result, there is no fixed phase correlation between neighboring points at the source, and the emitted beam is treated as spatially incoherent. After free-space propagation, however, spatial correlations appear in the downstream field. This behavior is a well-established result of classical wave optics \cite{van1934wahrscheinliche,zernike1938concept}.

For an extended incoherent or partially coherent source the Van Cittert-Zernike (VCZ) theorem states that the mutual coherence function in the observation plane is proportional to the scaled two-dimensional Fourier transform of the source intensity distribution. In the far-field (Fraunhofer) regime, the stationary mutual coherence function between two points  $(x_1,y_1)$ and $(x_2,y_2)$ in an observation plane located a distance z from the source, is given by:
\begin{equation}
\Gamma(\vec\Delta) \propto \iint I(\vec r') \,
e^{-i \frac{k}{z}\vec\Delta\cdot\vec r'} d^2\vec r'
\label{eq:integral_vcz}
\end{equation}
where $\vec\Delta=(x_2-x_1, y_2-y_1)$, $\vec r'=(x',y')$ denotes transverse coordinates in the source plane, $z$ is the propagation distance, $\lambda$ is the neutron wavelength, and $k=2\pi/\lambda$ is the neutron wavevector.

For illustration, consider a source of characteristic width $a$ with uniform intensity. The VCZ theorem then yields a mutual coherence function in the observation plane with a characteristic width set by the first zero of the resulting sinc profile. This defines a transverse coherence length of order
\begin{equation}
\zeta \approx \frac{\lambda z}{a} = \frac{\lambda}{\theta},
\label{Eqn:coherencelengthCZ}
\end{equation}
where $\theta \approx a/z$ characterizes the angular divergence of the beam. This inverse relationship between coherence length and angular divergence is a general consequence of the VCZ theorem and applies to arbitrary source distributions, including uniform, Gaussian, and other structured profiles.

 \begin{figure}[t]
    \centering\includegraphics[width=1\linewidth]{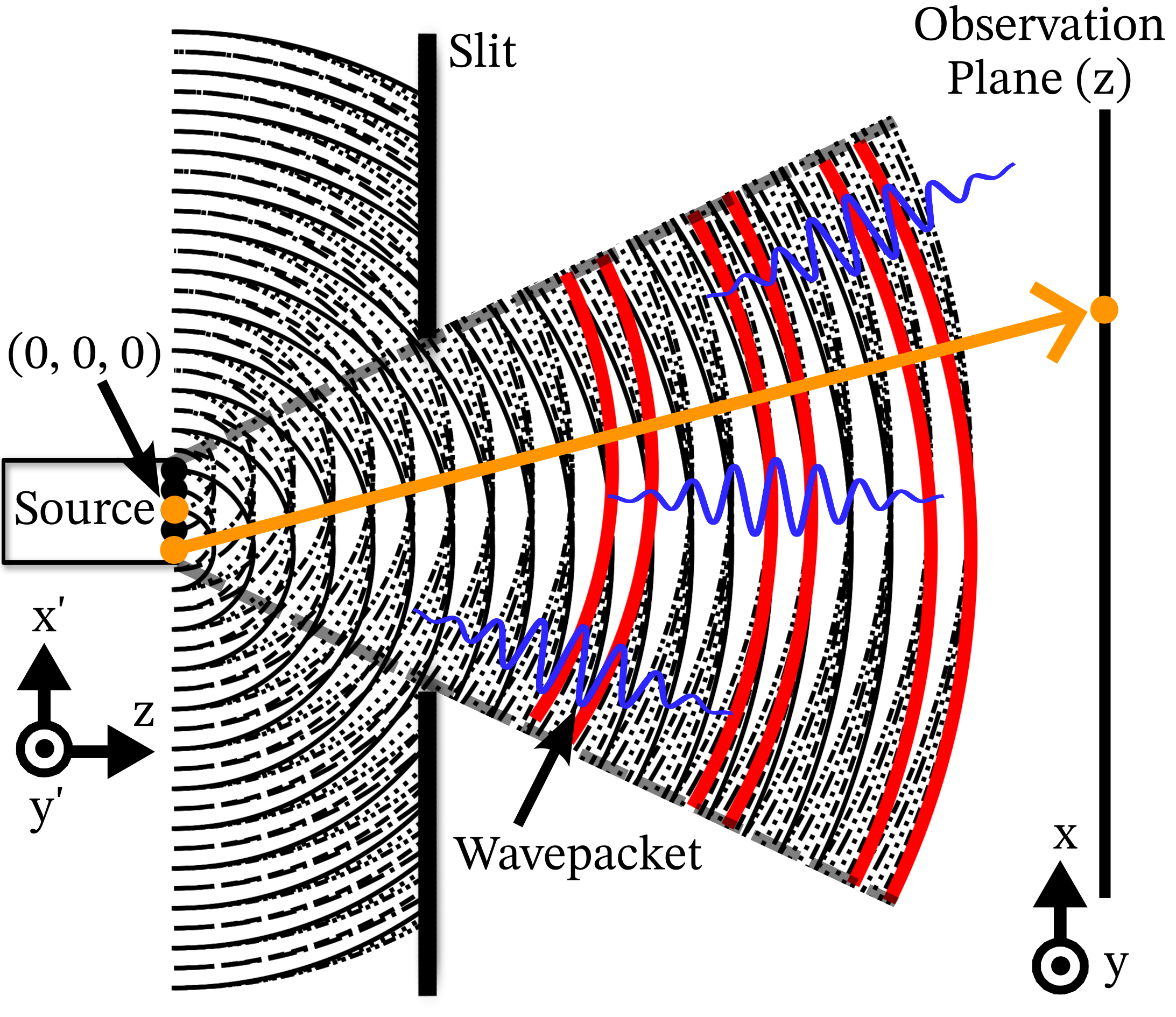}
    \caption{Illustration of two interpretations of transverse coherence in neutron beams. In the small-wavepacket picture, the coherence length is identified with the spatial extent of an individual neutron wavefunction. In the Van Cittert-Zernike description, coherence arises from the angular distribution of extended wavepackets emitted from different regions of an incoherent source, producing spatial correlations in the observation plane.}
    \label{fig:fig0}
\end{figure}

The physical origin of the VCZ theorem can be understood by decomposing the propagated field into tilted plane-wave components associated with different points of the source, as shown in Fig.~\ref{fig:fig0}. In this picture, the angular distribution of these plane waves determines the spatial correlations observed in the detection plane. Each point $\vec r'=(x',y')$ on the source emits a spherical wave
\begin{equation}
\psi(\vec r) = \frac{A(\vec r')}{\rho}e^{ik\rho},
\end{equation}
where $\vec r=(x,y)$ denotes transverse coordinates in the observation plane, $\rho$ is the distance between the source point and the observation point, and $A(\vec r')$ denotes the complex amplitude at the source. In the Fraunhofer regime and under the paraxial approximation:
\begin{equation}
\rho=\sqrt{(x-x')^2+(y-y')^2+z^2}.
\end{equation}

\noindent Expanding about the optical axis gives
\begin{equation}
\rho \approx z + \frac{x^2 + y^2}{2z} - \frac{xx' + yy'}{z} + \frac{x'^2 + y'^2}{2z}.
\end{equation}
The resulting phase term shows that each spherical wave is well approximated by a plane wave with a small transverse wavevector $\vec k_\perp = -k\,\vec r'/z$, corresponding to a propagation angle $\vec \theta \approx \vec r'/z$. The magnitude of this angle is $\theta \approx \sqrt{x'^2 + y'^2}/z$, demonstrating that different source points generate plane-wave components with different propagation angles.

The propagated field can therefore be expressed in the far field as a superposition of tilted plane-wave components originating from different points on the source:
\begin{equation}
\psi(\vec r)\approx \frac{e^{ikz}}{i\lambda z}A(\vec r')\,
e^{-i\frac{k}{ z}(xx'+yy')}.
\end{equation}

For a spatially incoherent source, fields originating from distinct points are statistically uncorrelated, satisfying
\begin{equation}
\left\langle
A^*(\vec r'_1)A(\vec r'_2)
\right\rangle
=
I(\vec r'_1)\,
\delta(\vec r'_1-\vec r'_2).
\end{equation}

Ensemble averaging then yields
\begin{equation}
\Gamma(\vec\Delta)
=
\iint
\langle |A(\vec r')|^2\rangle
\,e^{-i\frac{k}{z}\vec\Delta\cdot\vec r'}
\,d^2\vec r',
\end{equation}

Identifying $I(\vec r')=\langle|A(\vec r')|^2\rangle$ recovers the VCZ theorem.

\begin{figure}
    \centering\includegraphics[width=1\linewidth]{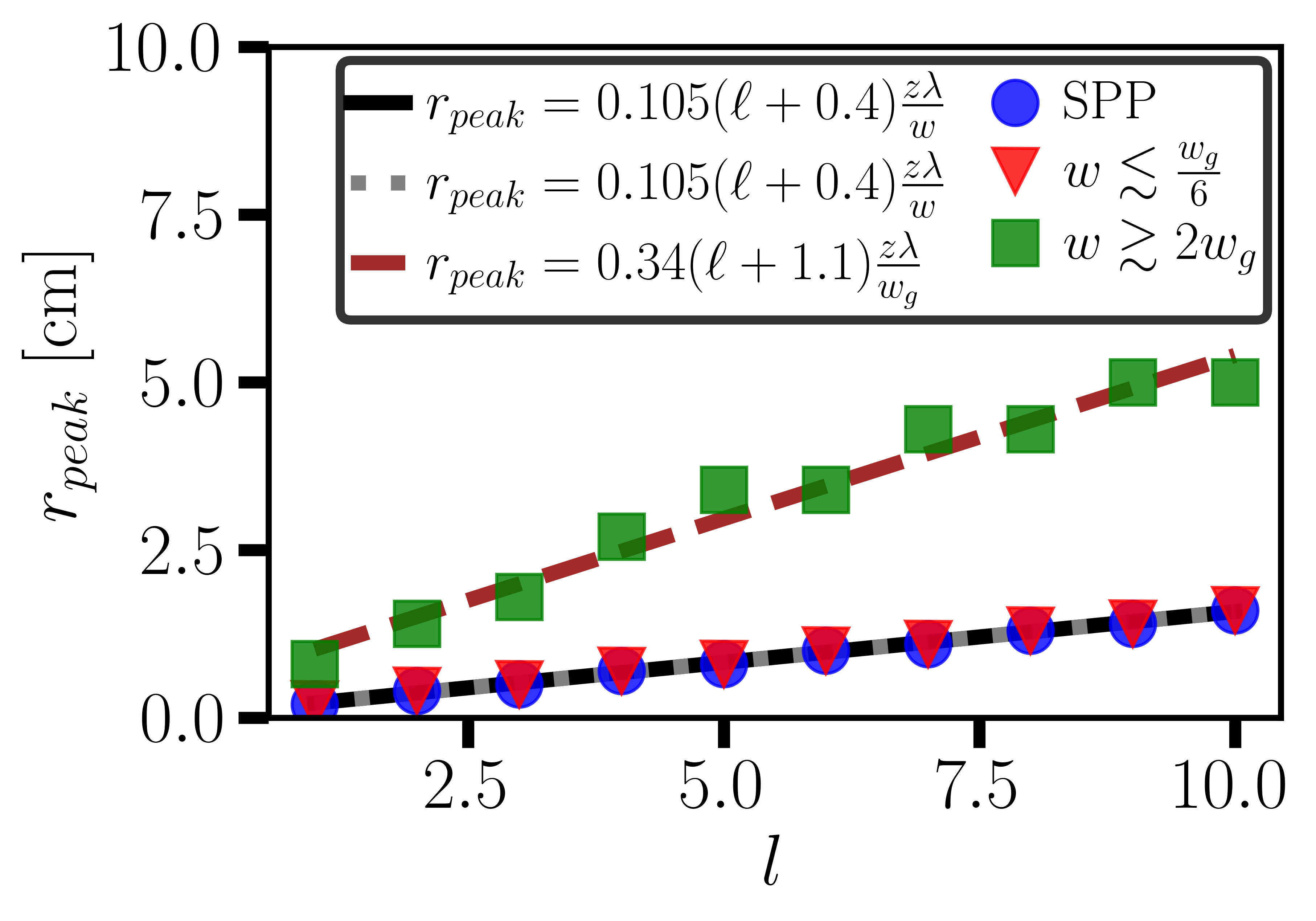}
    \caption{Relationship between helical mode $\ell$ and the radius of the intensity maximum $r_{peak}$ at propagation distance $z$, for a simulated Gaussian wavepacket, with transverse extent $w$, through 1. on-axis spiral phase plate (SPP), 2. fork dislocation phase grating with $w \lesssim w_g/6$, where $w_g$ is the grating width, 3. fork dislocation phase grating array, where the transverse extent follows $w \gtrsim 2w_g$. 
    The demonstrated simulation in 3. is for $z = 12$~m, $\lambda = 12$~\AA~, $w=3~\mu$m, and $w_g=1~\mu$m. The corresponding errors in the fitted parameters are: $0.105 \pm 0.002$, $0.4 \pm 0.1$, $0.34 \pm 0.02$, and $1.1 \pm 0.4$.}
    \label{fig:sim_methods}
\end{figure}

\subsection{Single-Neutron Description of Beam Coherence}

Neutron beams are often described as ensembles of plane waves with different wavevector magnitudes and directions, where the resulting distribution defines the instrument resolution. The previous sections showed how this description explains the observed transverse coherence through the VCZ theorem. We now examine how the same behavior arises when neutrons are detected individually.

After propagation through the beam-defining apertures, the transverse component of the neutron wavefunction can be written as a wavepacket
\begin{equation}
\psi(\vec r) = \int \tilde{\psi}(\vec k_\perp)\,
e^{i\vec k_\perp\cdot\vec r}\, d^2\vec k_\perp ,
\end{equation}
where $\vec r=(x,y)$ denotes transverse coordinates in the observation plane and $\tilde{\psi}(\vec k_\perp)$ is the transverse momentum-space amplitude centered around a mean propagation direction.

The beam itself can be modeled as an incoherent ensemble of such neutrons, each with a similar spatial envelope but a different mean transverse wavevector. The wavefunction for the $n$-th neutron can therefore be written as
\begin{equation}
\psi_n(\vec r)=\psi_0(\vec r)\,e^{i\vec k_{\perp,n}\cdot\vec r},
\end{equation}
where $\psi_0(\vec r)$ is a common envelope function of characteristic transverse width $w$, which defines the spatial extent of the individual neutron wavepacket, and $\vec k_{\perp,n}$ is drawn from the distribution $P(\vec k_\perp)$.

When neutrons are detected one at a time, the measured intensity is obtained by summing over many independent detection events:
\begin{equation}
I(\vec r)=\sum_n|\psi_n(\vec r)|^2
\approx\int P(\vec k_\perp)
\left|\psi_0(\vec r)e^{i\vec k_\perp\cdot\vec r}\right|^2
d^2\vec k_\perp .
\end{equation}
Since $|e^{i\vec k_\perp\cdot\vec r}|^2=1$, the phase factors cancel and, with $\int P(\vec k_\perp)\,d^2\vec k_\perp = 1$, the intensity reduces to
\begin{equation}
I(\vec r)=|\psi_0(\vec r)|^2 
\end{equation}
\noindent up to an overall normalization. To recover the mutual coherence function between two observation points
$\vec r_1=(x_1,y_1)$ and $\vec r_2=(x_2,y_2)$,
we compute the ensemble average
\begin{equation}
\begin{aligned}
\Gamma(\vec r_1,\vec r_2)
&=\left\langle\psi_n^*(\vec r_1)\psi_n(\vec r_2)\right\rangle_n \\
&=\psi_0^*(\vec r_1)\psi_0(\vec r_2)
\int P(\vec k_\perp)
e^{i\vec k_\perp\cdot(\vec r_2-\vec r_1)} d^2\vec k_\perp .
\end{aligned}
\end{equation}
For a Gaussian transverse momentum distribution of width $\Delta k_\perp$, this yields

\begin{equation}
\Gamma(\vec r_1,\vec r_2)\propto
\psi_0^*(\vec r_1)\psi_0(\vec r_2)
\exp\!\left(-\frac{|\vec r_1-\vec r_2|^2}{2\sigma^2}\right)
\end{equation}

The mutual coherence function contains contributions from both the single-particle envelope $\psi_0$ and the angular distribution of the ensemble. In the regime $w \gg \sigma$, the envelope varies only weakly over the spatial scale on which the Gaussian coherence term decays. Consequently, the characteristic width of the mutual coherence function is determined by $\sigma$ where:

\begin{equation}
\sigma=\frac{1}{\Delta k_\perp}=\frac{\lambda}{2\pi\theta}.
\label{Eq:sigmatoK}
\end{equation}

This has the same inverse dependence on angular divergence as the VCZ theorem result in Eq.~\ref{Eqn:coherencelengthCZ}, with numerical factors determined by the chosen source distribution and definition of coherence width. In practice, this coherence length is inferred from measurements of beam divergence, typically obtained using diffraction~\cite{shull1969single,treimer2006slit,wagh2011plain,altissimo2008neutron}, interferometry~\cite{pushin2008measurements}, or instrument resolution characterization ~\cite{majkrzak2022effect}.

\begin{figure}
    \centering\includegraphics[width=1\linewidth]{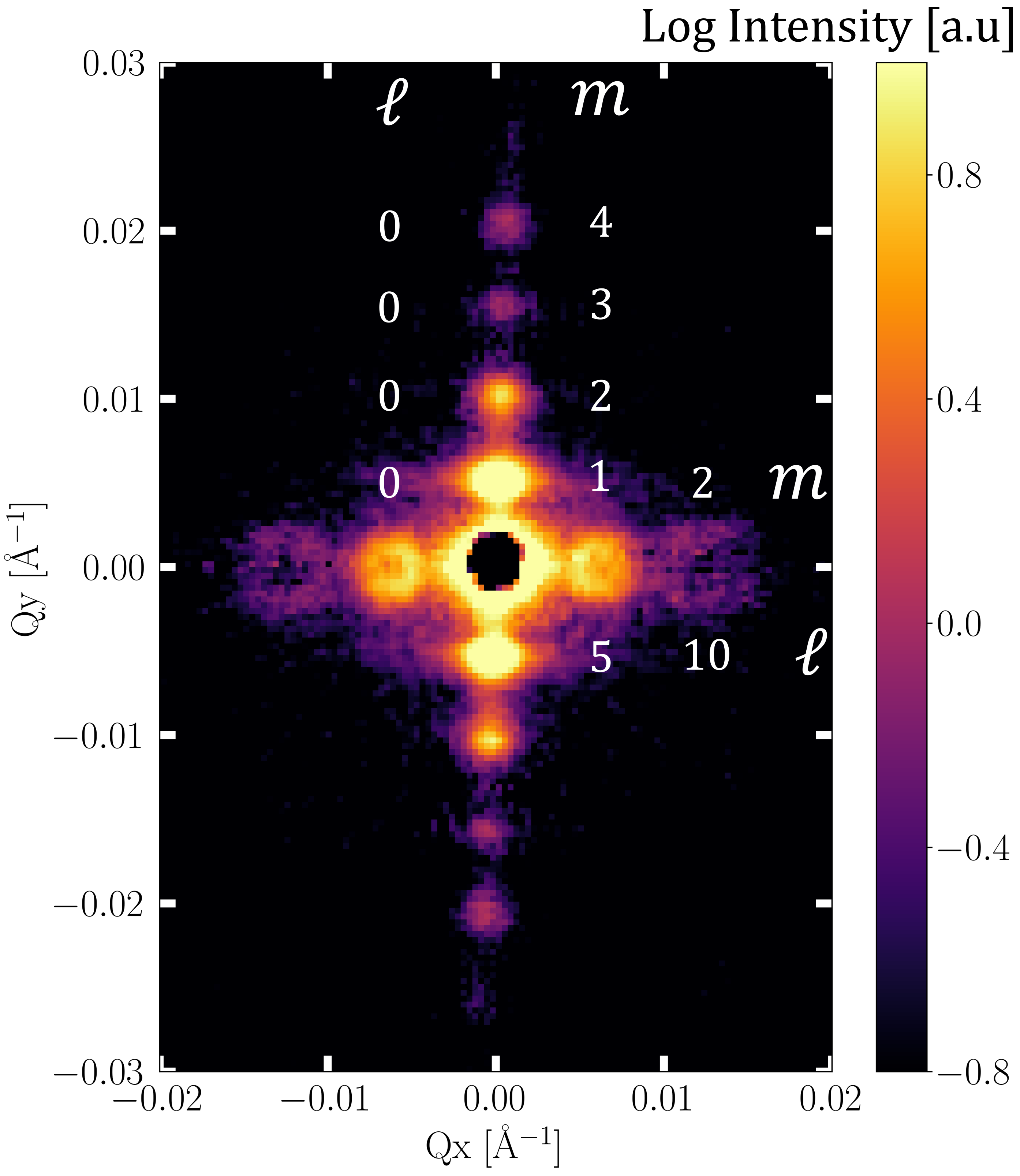}
    \caption{Experimental diffraction pattern showing $\ell = 0$ neutron states along the vertical axis and $\ell = 5 * m$ neutron states in the $m$ diffraction orders along the horizontal axis. The intensity is plotted on a logarithmic scale, revealing higher diffraction orders. The plotted wavelength range is $1.75-12.5~$\AA. Analogous measurements were taken for the $\ell = 3$ and $\ell = 7$ states shown in Fig.~\ref{fig:profs}.}
    \label{fig:oam_cross}
\end{figure}

\subsection{Decoupling Wavepacket Extent and Beam Coherence Using Helical Neutron States}

 \begin{figure*}
    \centering\includegraphics[width=1\linewidth]{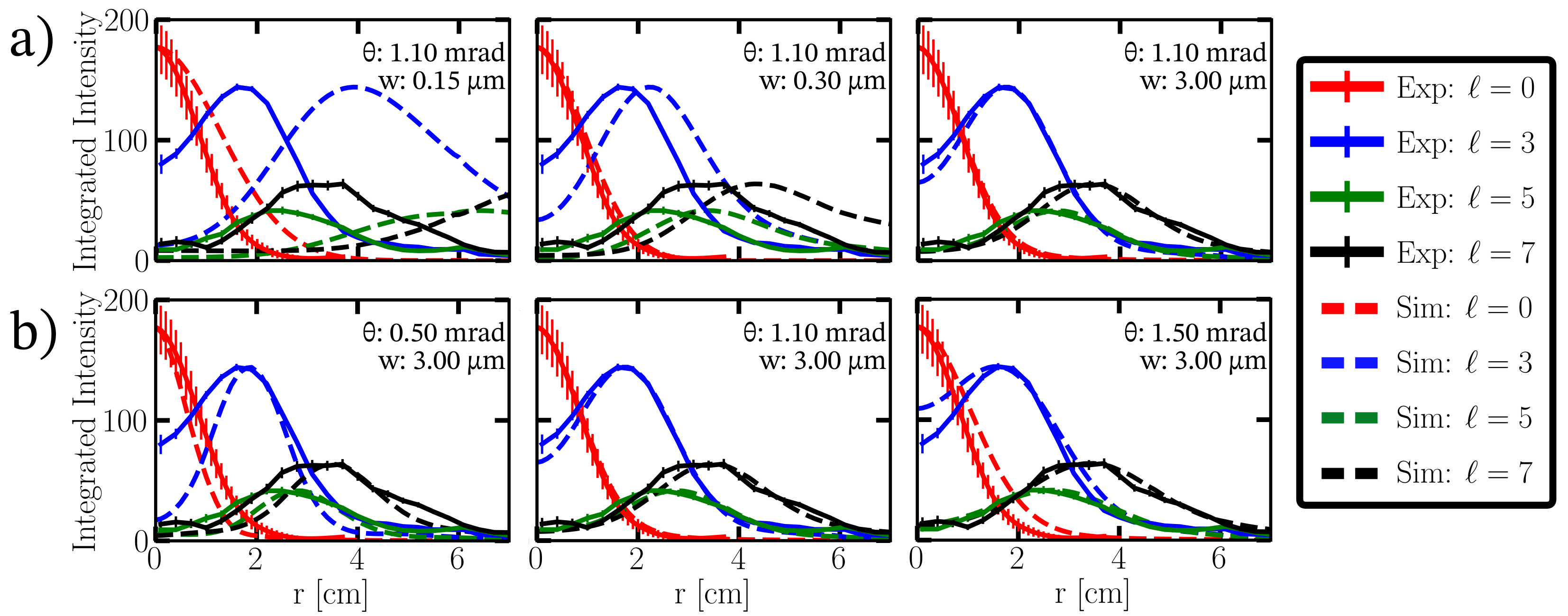}
    \caption{Experimental and simulated azimuthally integrated intensity profiles for the $\ell = 0, 3, 5, 7$ diffraction orders, for $11.5$~\AA~-$12.5$~\AA~neutrons. (a) The divergence angle is fixed at $\theta = 1.1$~mrad, while the input Gaussian width is varied for $w = 0.15~\mu$m, $w = 0.3~\mu$m, and $w = 3.00~\mu$m. The best agreement with experiment is obtained for $w \gtrsim 2~\mu$m, as summarized in Fig.~\ref{fig:profile_error}. (b) The input Gaussian width is fixed at $w = 3.00~\mu$m, while the divergence is varied for $\theta = 0.50$~mrad, $\theta = 1.1$~mrad, and $\theta = 1.50$~mrad. The optimal agreement is achieved for $\theta = 1.10$~mrad, as summarized in Fig.~\ref{fig:profile_error}. Experimental profiles are spatially binned to an effective pixel size of $\leq$ 4 mm to improve the robustness of the fits to the simulated profiles. Uncertainties shown are statistical.}
    % 3, 4, 5, 6, 8 mm convolution sigmas = 0.5, 0.67, 0.83, 1.0, 1.33 mrad
    \label{fig:profs}
\end{figure*}

Here we show that helical neutron states that carry orbital angular momentum (OAM) provide a direct experimental probe of the transverse structure of the neutron wavefunction. In conventional measurements, both the transverse coherence length $\sigma$ and the spatial extent of the wavepacket contribute to the overall width of the observed intensity distribution, making them difficult to distinguish. In contrast, helical beams exhibit a characteristic annular intensity profile whose peak radius depends explicitly on the transverse wavepacket extent. 

The neutron wavefunction corresponding to the first diffraction order immediately after the fork-dislocation phase grating is well approximated by a helically phased Gaussian of the form

\begin{equation}
\psi(r,\phi) = \psi_0(r)e^{i\ell\phi}
= \sum_p c_p \, \mathrm{LG}_{p,\ell}(r,\phi),
\label{eq:sppState}
\end{equation}

where $c_p$ are expansion coefficients and the field is expressed as a superposition of Laguerre--Gaussian (LG) modes with fixed topological charge $\ell$ and a distribution of radial indices $p$. For a pure $p=0$ LG mode, the radius of maximum intensity scales as $\sqrt{|\ell|}$~\cite{padgett2015divergence}. On the other hand, for a state with a superposition over radial modes (Eq.~\ref{eq:sppState}), the peak radius exhibits an approximately linear dependence on $|\ell|$~\cite{padgett2015divergence}. This relationship between helical mode $\ell$ and the radius of the intensity maximum $r_{peak}$ at propagation distance $z$, can be characterized by numerically propagating Eq.~\ref{eq:sppState} and performing a least-squares fit to the simulated intensity profiles maxima to obtain:

\begin{equation}
r_{\mathrm{peak}} = 0.11(\ell+0.4)\frac{\lambda z}{w},
\end{equation}
where $w$ denotes the transverse extent of the neutron wavepacket. %In the limit of large transverse coherence and a single phase element, this dependence would enable a direct determination of $w$ from the peak position. 

In the present configuration, however, the use of a phase-grating array together with finite beam coherence modifies the observed profiles, so that the measurement provides a lower bound on $w$ rather than a precise value. Notably, for a wavepacket propagating through a fork dislocation phase-grating array, the physical size of the grating $w_g$ relative to $w$ modifies this behaviour (see Fig.~\ref{fig:sim_methods}), yielding:
\begin{equation}
r_{\mathrm{peak}} = 0.34(\ell+1.1)\frac{\lambda z}{w_g},
\end{equation}

The finite coherence length broadens the observed profile but does not determine the location of this peak. Consequently, measuring the radial position of the OAM intensity maximum provides direct sensitivity to the wavepacket extent $w$, independently of the transverse coherence length $\sigma$ that characterizes the ensemble divergence of the beam.

\section{Methods}
\label{oam_methods}
We performed small-angle neutron scattering (SANS) measurements on the SANS2D instrument at the ISIS Neutron and Muon Source~\cite{heenan2011small}. The instrument geometry was: source aperture $A_1 = 20\times20$~mm, collimation length $L_1 = 12$~m, sample aperture $A_2 = 6\times6$~mm, and sample-to-detector distance $L_2 = 12$~m. The geometric angular divergence is therefore estimated as:
\begin{equation}
\theta \approx \frac{A_1/2 + A_2/2}{L_1}
=1.08~\mathrm{mrad}.
\end{equation}

The usable wavelength range at this configuration was $1.75$--$12.5$~\AA, with a spectral peak near $3$~\AA. All data were reduced using the standard Mantid framework for ISIS instruments~\cite{arnold2014mantid}.

We implemented four fork-dislocation phase gratings that generate neutron orbital angular momentum (OAM) states with $\ell = 0, 3, 5, 7$ in the first diffraction order. The phase gratings consisted of more than six million micron-scale phase elements etched into a silicon wafer, with linear grating periods of $p=120$~nm for $\ell = 0,3,7$ and $p=100$~nm for $\ell = 5$. These gratings have been described in detail in Ref.~\cite{sarenac2022experimental}. 

\section{Results and Discussion}

We experimentally generated neutron helical states with OAM numbers $\ell = 0, 3, 5, 7$, providing an experimental platform for separating the effects of beam divergence from the transverse extent of the neutron wavepacket. The experimental diffraction pattern for $\ell = 0$ and $\ell =5$ across the entire wavelength range is shown in Fig.~\ref{fig:oam_cross}. 

Shown in Fig.~\ref{fig:profs} are the azimuthally integrated experimental intensity profiles for $\ell = 0, 3, 5, 7$, alongside simulation. In Fig.~\ref{fig:profs}a, the simulated divergence angle is fixed at the value of $1.1$~mrad and the input Gaussian width $w$ in the simulation is varied. Agreement with experiment improves as the assumed wavepacket width $w$ increases. In Fig.~\ref{fig:profs}b the input Gaussian width $w$ is fixed at $3~\mu$m and the simulated divergence angle is varied.

\begin{figure}
    \centering\includegraphics[width=1\linewidth]{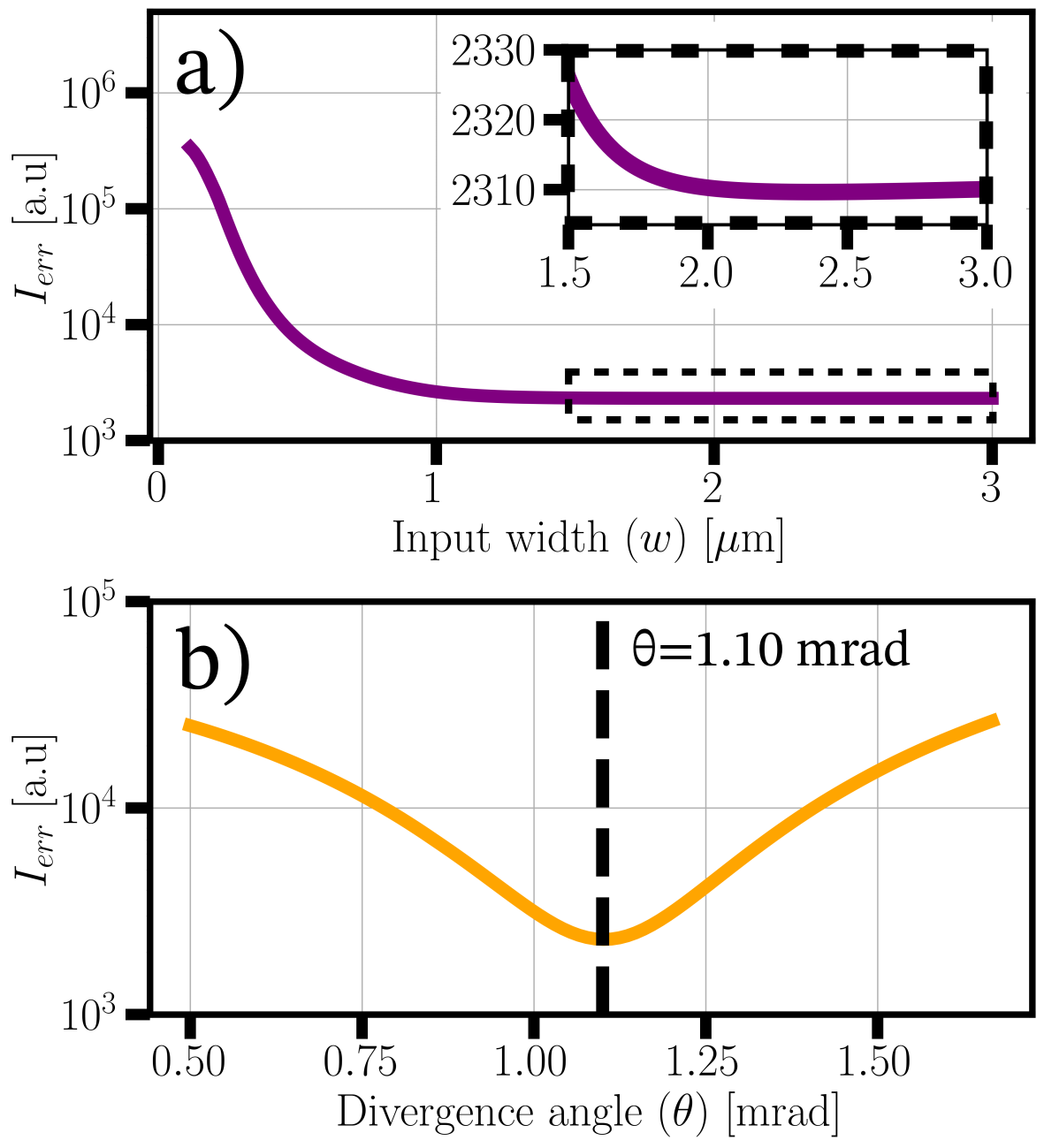}
    \caption{(a) Error between simulation and experiment for the azimuthally integrated OAM intensity profiles (see Fig.~\ref{fig:profs}) for $\ell = 0, 3, 5, 7$ as a function of the simulated input Gaussian width, with the divergence angle fixed at $\theta = 1.1$~mrad. For input widths $\gtrsim 2~\mu$m, the simulation shows good agreement with experiment, as demonstrated in the inset. (b) Error between simulation and experiment for the azimuthally integrated OAM intensity profiles (see Fig.~\ref{fig:profs}) for $\ell = 0, 3, 5, 7$ as a function of the divergence angle $\theta$, with the input Gaussian width fixed at $w = 3~\mu$m. The optimal divergence is $\theta = 1.1$~mrad.}
    \label{fig:profile_error}
\end{figure}

This behaviour is characterized by computing the total error between experimental and simulated intensity profiles:
\begin{equation}
    I_{err} = \sum_{\ell = 0,3,5,7} \sum_r \left[I_{sim}(r, \ell, \theta, w) - I_{exp}(r, \ell)\right]^2, 
\end{equation}
where $I_{sim}(r, \ell, \theta, w)$ and $ I_{exp}(r, \ell)$ are the simulated and experimental azimuthally integrated intensity profiles shown in Fig.~\ref{fig:profs}. Shown in Fig.~\ref{fig:profile_error} is a plot of $I_{err}$ where we independently vary $w$ and $\theta$. The upstream collimation optics determine the angular divergence of the beam, which for the present geometry is approximately:
\begin{equation}
\theta = 1.1~\mathrm{mrad}.
\end{equation}
The optimal divergence angle obtained from the fit agrees with the independently determined experimental divergence, while the optimal wavepacket extent is greater than $2~\mu$m. Using Eq.~\ref{Eq:sigmatoK} this corresponds to a transverse coherence length of $\sigma \approx 180~\mathrm{nm}$ for $\lambda = 12~\text{\AA}$.

As the transverse coherence length decreases, the characteristic annular structure of the OAM intensity profile progressively evolves toward a Gaussian-like distribution, as illustrated in Fig.~\ref{fig:profs}. While the OAM peak position remains identifiable, coherence-induced broadening increasingly influences the observed profile and can shift the fitted intensity maximum. Consequently, measurements performed while the annular OAM structure remains resolvable provide a lower bound on the wavepacket extent, whereas for sufficiently poor coherence the method loses sensitivity to the wavepacket size altogether.

\section{Conclusion}

We have introduced and experimentally demonstrated a method for separating transverse coherence length from single-particle wavepacket extent using helical neutron wavefronts. Unlike conventional measurements, in which both quantities contribute to spatial broadening, the annular intensity profiles of helical neutron states provide an observable whose peak position depends on wavepacket extent while coherence primarily contributes to profile broadening.

From the measured beam divergence of $\theta \approx 1.1~\mathrm{mrad}$ we obtain a transverse coherence length of approximately $180~\mathrm{nm}$. In contrast, analysis of the helical intensity profiles establishes a lower bound of $w \ge 2~\mu\mathrm{m}$ on the spatial extent of the individual neutron wavepacket. The wavepacket extent therefore exceeds the transverse coherence length by more than an order of magnitude. Such a disparity is not unexpected, as the transverse extent of a freely propagating neutron wavepacket can become comparable to the beam dimensions through Schrödinger evolution, whereas the transverse coherence length remains determined by the ensemble angular distribution.

These results provide direct experimental evidence that transverse coherence length and single-particle wavepacket extent are distinct physical quantities. As a consequence, observations of increased coherence following apertures, collimation, or other post-selection procedures do not imply a corresponding increase in the size of the individual neutron wavepacket. Although the present analysis employs Gaussian wavepackets and Gaussian angular distributions for clarity, the distinction between ensemble coherence and single-particle extent does not depend on these assumptions. Alternative source profiles governed by the VCZ theorem, as well as non-Gaussian wavepacket structures such as Airy or other structured states~\cite{sarenac2025generation,geerits2023phase,McKay2026Topological}, preserve the separation between ensemble coherence and individual-particle spatial extent. Finally, the present work focuses on transverse degrees of freedom; extending similar approaches to investigate the relationship between longitudinal coherence length and longitudinal wavepacket extent remains an important direction for future study.

\section{Acknowledgments}
This work was supported by the Canadian Excellence Research Chairs (CERC) program, the Natural Sciences and Engineering Research Council of Canada (NSERC) Discovery program, the NSERC Canada Graduate Scholarships programs (CGS-M and PGS-D), the Collaborative Research and Training Experience (CREATE) program, the Canada  First  Research  Excellence  Fund  (CFREF), and the US Department of Energy, Office of Nuclear Physics, under Interagency Agreement 89243019SSC000025. This work was also supported by the DOE Office of Science, Office of Basic Energy Sciences, in the program “Quantum Horizons: QIS Research and Innovation for Nuclear Science” through grant DE-SC0023695. Experiments at the ISIS Neutron and Muon Source were supported by beamtime allocation RB2510543 from the Science and Technology Facilities Council.

\bibliographystyle{ieeetr}
\bibliography{refs}

@article{McKay2026Topological,
  title = {Topological shaping of vortex neutron beams using forked phase gratings},
  author = {McKay, S. and Parnell, S. R. and Dalgliesh, R. M. and Lavrik, N. V. and Kravchenko, I. I. and Le Thien, Q. and Baxter, D. V. and Ortiz, G. and Pynn, R.},
  journal = {Phys. Rev. Res.},
  volume = {8},
  issue = {2},
  pages = {023199},
  numpages = {7},
  year = {2026},
  month = {May},
  publisher = {American Physical Society},
  doi = {10.1103/prnf-wtjp},
  url = {https://link.aps.org/doi/10.1103/prnf-wtjp}
}

@article{geerits2023phase,
  title={Phase vortex lattices in neutron interferometry},
  author={Geerits, Niels and Lemmel, Hartmut and Berger, Anna-Sophie and Sponar, Stephan},
  journal={Communications Physics},
  volume={6},
  number={1},
  pages={209},
  year={2023},
  publisher={Nature Publishing Group UK London}
}

@article{altissimo2008neutron,
  title={Neutron diffraction from macroscopic objects and transverse coherence of the wavefunction: The Fresnel zone plates},
  author={Altissimo, Matteo and Petrillo, Caterina and Sacchetti, Francesco and Sani, Lorenzo and Stahn, Jochen},
  journal={Nuclear Instruments and Methods in Physics Research Section A: Accelerators, Spectrometers, Detectors and Associated Equipment},
  volume={586},
  number={1},
  pages={68--72},
  year={2008},
  publisher={Elsevier}
}

@article{sarenac2025generation,
  title={Generation of neutron Airy beams},
  author={Sarenac, Dusan and Lailey, Owen and Henderson, Melissa E and Ekinci, Huseyin and Clark, Charles W and Cory, David G and DeBeer-Schmitt, Lisa and Huber, Michael G and White, Jonathan S and Zhernenkov, Kirill and others},
  journal={Physical Review Letters},
  volume={134},
  number={15},
  pages={153401},
  year={2025},
  publisher={APS}
}

@article{fischer2006neutron,
  title={Neutron and x-ray diffraction studies of liquids and glasses},
  author={Fischer, Henry E and Barnes, Adrian C and Salmon, Philip S},
  journal={Reports on Progress in Physics},
  volume={69},
  number={1},
  pages={233--299},
  year={2006}
}

@article{lailey2025multimode,
  title={Multimode structured neutron beams},
  author={Lailey, Owen and Sarenac, Dusan and Clark, Charles W and Cory, David G and DeBeer-Schmitt, Lisa and Ekinci, Huseyin and Garrad, Davis V and Henderson, Melissa E and Huber, Michael G and Vadnere, Priyanka and others},
  journal={arXiv preprint arXiv:2511.07662},
  year={2025}
}

@article{sarenac2026experimental,
  title={Experimental Limit on Neutron Orbital Angular Momentum Detection Using Polarized 3He},
  author={Sarenac, D and Lailey, O and Garrad, DV and Vadnere, PR and Shentevski, N and Clark, CW and Cory, DG and Cotter, JP and Ekinci, H and Huber, MG and others},
  journal={arXiv preprint arXiv:2603.16655},
  year={2026}
}

@article{padgett2015divergence,
  title={Divergence of an orbital-angular-momentum-carrying beam upon propagation},
  author={Padgett, Miles J and Miatto, Filippo M and Lavery, Martin PJ and Zeilinger, Anton and Boyd, Robert W},
  journal={New Journal of Physics},
  volume={17},
  number={2},
  pages={023011},
  year={2015},
  publisher={IOP Publishing}
}

@article{sarenac2024small,
  title={Small-angle scattering interferometry with neutron orbital angular momentum states},
  author={Sarenac, Dusan and Henderson, Melissa E and Ekinci, Huseyin and Clark, Charles W and Cory, David G and DeBeer-Schmitt, Lisa and Huber, Michael G and Lailey, Owen and White, Jonathan S and Zhernenkov, Kirill and others},
  journal={Nature Communications},
  volume={15},
  number={1},
  pages={10785},
  year={2024},
  publisher={Nature Publishing Group UK London}
}

@article{pushin2008measurements,
  title={Measurements of the vertical coherence length in neutron interferometry},
  author={Pushin, DA and Arif, M and Huber, MG and Cory, DG},
  journal={Physical review letters},
  volume={100},
  number={25},
  pages={250404},
  year={2008},
  publisher={APS}
}

@article{wagh2011plain,
  title={The plain truth about forming a plane wave of neutrons},
  author={Wagh, Apoorva G and Abbas, Sohrab and Treimer, Wolfgang},
  journal={Nuclear Instruments and Methods in Physics Research Section A: Accelerators, Spectrometers, Detectors and Associated Equipment},
  volume={634},
  number={1},
  pages={S41--S45},
  year={2011},
  publisher={Elsevier}
}

@article{treimer2006slit,
  title={Slit and phase grating diffraction with a double crystal diffractometer},
  author={Treimer, Wolfgang and Hilger, Andr{\'e} and Strobl, Markus},
  journal={Physica B: Condensed Matter},
  volume={385},
  pages={1388--1391},
  year={2006},
  publisher={Elsevier}
}

@article{majkrzak2014determination,
  title={Determination of the effective transverse coherence of the neutron wave packet as employed in reflectivity investigations of condensed-matter structures. I. Measurements},
  author={Majkrzak, Charles F and Metting, Christopher and Maranville, Brian B and Dura, Joseph A and Satija, Sushil and Udovic, Terrence and Berk, Norman F},
  journal={Physical Review A},
  volume={89},
  number={3},
  pages={033851},
  year={2014},
  publisher={APS}
}

@article{zernike1938concept,
  title={The concept of degree of coherence and its application to optical problems},
  author={Zernike, Frederik},
  journal={Physica},
  volume={5},
  number={8},
  pages={785--795},
  year={1938},
  publisher={Elsevier}
}

@article{van1934wahrscheinliche,
  title={Die wahrscheinliche Schwingungsverteilung in einer von einer Lichtquelle direkt oder mittels einer Linse beleuchteten Ebene},
  author={van Cittert, Pieter Hendrik},
  journal={Physica},
  volume={1},
  number={1-6},
  pages={201--210},
  year={1934},
  publisher={Elsevier}
}

@article{glauber1963coherent,
  title={Coherent and incoherent states of the radiation field},
  author={Glauber, Roy J},
  journal={Physical Review},
  volume={131},
  number={6},
  pages={2766},
  year={1963},
  publisher={APS}
}

@article{glauber1963quantum,
  title={The quantum theory of optical coherence},
  author={Glauber, Roy J},
  journal={Physical Review},
  volume={130},
  number={6},
  pages={2529},
  year={1963},
  publisher={APS}
}

@book{utsuro2010handbook,
  title={Handbook of neutron optics},
  author={Utsuro, Masahiko and Ignatovich, Vladimir K},
  year={2010},
  publisher={John Wiley \& Sons}
}

@article{majkrzak2022effect,
  title={The effect of transverse wavefront width on specular neutron reflection},
  author={Majkrzak, CF and Berk, Norman F and Maranville, Brian B and Dura, JA and Jach, T},
  journal={Applied Crystallography},
  volume={55},
  number={4},
  pages={787--812},
  year={2022},
  publisher={International Union of Crystallography}
}

@article{shull1969single,
  title={Single-slit diffraction of neutrons},
  author={Shull, CG},
  journal={Physical Review},
  volume={179},
  number={3},
  pages={752},
  year={1969},
  publisher={APS}
}

@article{jach2022method,
  title={Method for the definitive detection of orbital angular momentum states in neutrons by spin-polarized he 3},
  author={Jach, Terrence and Vinson, John},
  journal={Physical Review C},
  volume={105},
  number={6},
  pages={L061601},
  year={2022},
  publisher={APS}
}

@article{treimer2025neutron,
  title={On neutron holography, neutron interferometry, and neutron orbital angular momentum},
  author={Treimer, Wolfgang and Hausser, Frank and Beckers, Ingeborg and Suda, Martin and Jach, Terrence},
  journal={Optics Express},
  volume={33},
  number={20},
  pages={42385--42401},
  year={2025},
  publisher={Optica Publishing Group}
}

@article{treimer2024computerized,
  title={Computerized simulation of 2-dimensional phase contrast images using spiral phase plates in neutron interferometry},
  author={Treimer, Wolfgang and Hau{\ss}er, Frank and Suda, Martin},
  journal={Zeitschrift f{\"u}r Naturforschung A},
  volume={79},
  number={9},
  pages={873--880},
  year={2024},
  publisher={De Gruyter}
}

@article{cappelletti2018intrinsic,
  title={Intrinsic orbital angular momentum states of neutrons},
  author={Cappelletti, Ronald L and Jach, Terrence and Vinson, John},
  journal={Physical review letters},
  volume={120},
  number={9},
  pages={090402},
  year={2018},
  publisher={APS}
}

@book{rauch2015neutron,
  title={Neutron interferometry: lessons in experimental quantum mechanics},
  author={Rauch, Helmut and Werner, Samuel A},
  year={2015},
  publisher={Oxford University Press, New York}
}

@article{klein1983longitudinal,
  title={Longitudinal coherence in neutron interferometry},
  author={Klein, AG and Opat, G I\_ and Hamilton, WA},
  journal={Physical Review Letters},
  volume={50},
  number={8},
  pages={563},
  year={1983},
  publisher={APS}
}

@article{heenan2011small,
  title={Small angle neutron scattering using Sans2d},
  author={Heenan, RK and Rogers, SE and Turner, D and Terry, AE and Treadgold, J and King, SM},
  journal={Neutron News},
  volume={22},
  number={2},
  pages={19--21},
  year={2011},
  publisher={Taylor \& Francis}
}

@article{arnold2014mantid,
  title={Mantid—Data analysis and visualization package for neutron scattering and $\mu$ SR experiments},
  author={Arnold, Owen and Bilheux, Jean-Christophe and Borreguero, JM and Buts, Alex and Campbell, Stuart I and Chapon, L and Doucet, Mathieu and Draper, N and Leal, R Ferraz and Gigg, MA and others},
  journal={Nuclear instruments and methods in physics research section a: accelerators, spectrometers, detectors and associated equipment},
  volume={764},
  pages={156--166},
  year={2014},
  publisher={Elsevier}
}

@article{sarenac2022experimental,
  title={Experimental realization of neutron helical waves},
  author={Sarenac, Dusan and Henderson, Melissa E and Ekinci, Huseyin and Clark, Charles W and Cory, David G and DeBeer-Schmitt, Lisa and Huber, Michael G and Kapahi, Connor and Pushin, Dmitry A},
  journal={Science Advances},
  volume={8},
  number={46},
  pages={eadd2002},
  year={2022},
  publisher={American Association for the Advancement of Science}
}

@Article{Kaiser1983,
  author    = {Kaiser, H. and Werner, S. A. and George, E. A.},
  journal   = {Physical Review Letters},
  title     = {Direct Measurement of the Longitudinal Coherence Length of a Thermal Neutron Beam},
  year      = {1983},
  month     = {Feb},
  number    = {8},
  pages     = {560--563},
  volume    = {50},
  doi       = {10.1103/PhysRevLett.50.560},
  numpages  = {3},
  publisher = {American Physical Society},
}

@Article{Rauch1996,
  author    = {Rauch, H. and W\"olwitsch, H. and Kaiser, H. and Clothier, R. and Werner, S. A.},
  journal   = {Physical Review},
  title     = {Measurement and characterization of the three-dimensional coherence function in neutron interferometry},
  year      = {1996},
  month     = {Feb},
  number    = {2},
  pages     = {902--908},
  volume    = {A53},
  doi       = {10.1103/PhysRevA.53.902},
  numpages  = {6},
  publisher = {American Physical Society},
}

@Article{Hamilton1983,
  author    = {Hamilton, W. A. and Klein, A. G. and Opat, G. I.},
  journal   = {Physical Review},
  title     = {Longitudinal coherence and interferometry in dispersive media},
  year      = {1983},
  month     = {Nov},
  number    = {5},
  pages     = {3149--3152},
  volume    = {A28},
  doi       = {10.1103/PhysRevA.28.3149},
  numpages  = {3},
  publisher = {American Physical Society},
}

@article{kardjilov2011neutron,
  title={Neutron imaging in materials science},
  author={Kardjilov, Nikolay and Manke, Ingo and Hilger, Andr{\'e} and Strobl, Markus and Banhart, John},
  journal={Materials Today},
  volume={14},
  number={6},
  pages={248--256},
  year={2011},
  publisher={Elsevier}
}

@book{lindner2024neutrons,
  title={Neutrons, X-rays, and light: scattering methods applied to soft condensed matter},
  author={Lindner, Peter and Oberdisse, Julian},
  year={2024},
  publisher={Elsevier}
}

\end{document}